\documentclass[twoside]{dis09}
\usepackage[latin1]{inputenc}
\usepackage[dvips]{graphicx,epsfig,color}
\usepackage{wrapfig,rotating}
\usepackage{amssymb,amsmath,array}

\newcommand{\stdfig}[4]{
  \begin{figure}[tbh]
  \centering
  \includegraphics[width=#2]{#1}
  \caption[#4]{#3}
  \label{fig:#1}
  \end{figure}
}

\newcommand{\gevsq}{\,\mathrm{GeV}^2}
\newcommand{\pbi}{\,\mathrm{pb}^{-1}}

\newcommand{\pth}{p_{t, HFS}}
\newcommand{\pte}{p_{t, e}}
\newcommand{\empz}{E-p_z}

\def\dof{\mathop{n_{\rm dof}}\nolimits}

\begin{document}
\title{Measurement of $\mathbf{F_2}$ at Medium $\mathbf{Q^2}$
  and the PDF Determination using H1 HERA I Data}

\author{Jan Kretzschmar
%
%
\vspace{.3cm}\\
%
University of Liverpool - Department of High Energy Physics\\
Oliver Lodge Laboratory, Liverpool L69 7ZE, UK
}

\maketitle

\begin{abstract}
  A new measurement of the inclusive DIS cross section $ep \rightarrow
  e'X$ using H1 data from the year 2000 is performed and combined with
  published results using 1996/97 data. The new data covers the region
  of $12\gevsq \leq Q^2 \leq 150\gevsq$ at HERA with unprecedented
  accuracy. The structure Function $F_2$ is extracted and studied.
  A QCD analysis, H1PDF~2009, is performed including the
  new and published H1 inclusive measurements.
\end{abstract}

\section{Introduction}

The HERA collider facility in Hamburg, Germany, was a unique machine
for lepton-proton scattering at highest energies. Here a new analysis
of data taken with the H1 detector in the year 2000 is
discussed~\cite{h1medq2}, when protons with an energy of $920$~GeV
were collided with positrons with an energy of $27.6$~GeV,
corresponding to a centre of mass energy of $\sqrt{s} = 320$~GeV.

Deep Inelastic Scattering (DIS) of leptons off nucleons continues
to be the tool for high precision measurements of the quark and gluon
content of the nucleons in the form of so called parton distribution
functions (PDFs). The evolution of PDFs is a sensitive test of our
understanding of QCD dynamics, which is expressed in the form of
evolution equations. Furthermore a precise knowledge of PDFs is vital
for measurements at hadron colliders, such as the LHC.

The kinematics of the scattering are described in terms of the Lorentz
invariant quantities: the Bjorken scaling variable $x$, the
inelasticity $y$, and the virtuality $Q^2$. The neutral
current (NC) inclusive cross section for the reaction $ep \rightarrow
e'X$ at low $Q^2 \ll M_Z^2$ can be expressed in the form
\begin{equation} \label{sigmar}
  \frac{\mathrm{d}^2\sigma^{NC}}{\mathrm{d}x\mathrm{d}Q^2} 
  = \frac{2\pi\alpha^2Y_+}{xQ^4}
  \left(F_2(x, Q^2) -
    \frac{y^2}{Y_+} F_L(x, Q^2) \right)
\end{equation}
with $Y_+ = 1 + (1-y)^2$ and the structure functions $F_2$ and $F_L$.
At leading order, the structure function $F_{2}$ relates to the quark
distribution functions simply as
\begin{equation}
  F_2 = x\sum e_q^2 (q(x) + \bar{q}(x)) \,.
\end{equation}
In higher orders the Bjorken scaling is violated and $F_2$ depends
also on $Q^2$. The derivative $\partial F_2(x, Q^2) /\partial \ln Q^2$
is related to the gluon distribution $xg(x, Q^2)$ and the value of the
strong coupling constant $\alpha_s$.

The longitudinal structure function vanishes at leading order. Its
contribution to the cross section is $\propto y^2$ and thus
very small for the new analysis, which is restricted to $y<0.6$.
Analyses focusing on the high $y$ domain and the $F_L$ determination
have been presented elsewhere~\cite{h1zeusfl}.

\section{New Cross Section Measurement}

The new measurement of the DIS cross section is performed using data
taken with the H1 apparatus in the year 2000. The integrated
luminosity of the sample is $22\pbi$. For $12 \gevsq
\leq Q^2 \leq 150 \gevsq$ the scattered positron is detected in the
backward region of H1. The SpaCal calorimeter and the planar drift
chamber BDC are used to identify the scattered positron and to measure
its energy $E_e'$. The Central Tracker is used to reconstruct the
event vertex and, in connection with the backward detectors, the
positron scattering angle $\theta_e$. The hadronic final state (HFS)
is reconstructed from a combination of tracks and calorimeter
measurements, where the transverse momentum, $P_T^h$, and the total
difference between energy and longitudinal momentum $\empz = \sum_i
(E_i - P_{z,i}) + E_e'(1-\cos \theta_e)$ are especially useful
quantities. Control distributions of these four basic variables
characterising the scattered positron and the HFS are shown in figure
\ref{fig:kretzschmar_jan.fig1.eps}, where an excellent agreement
between data and simulation is observed.

\stdfig{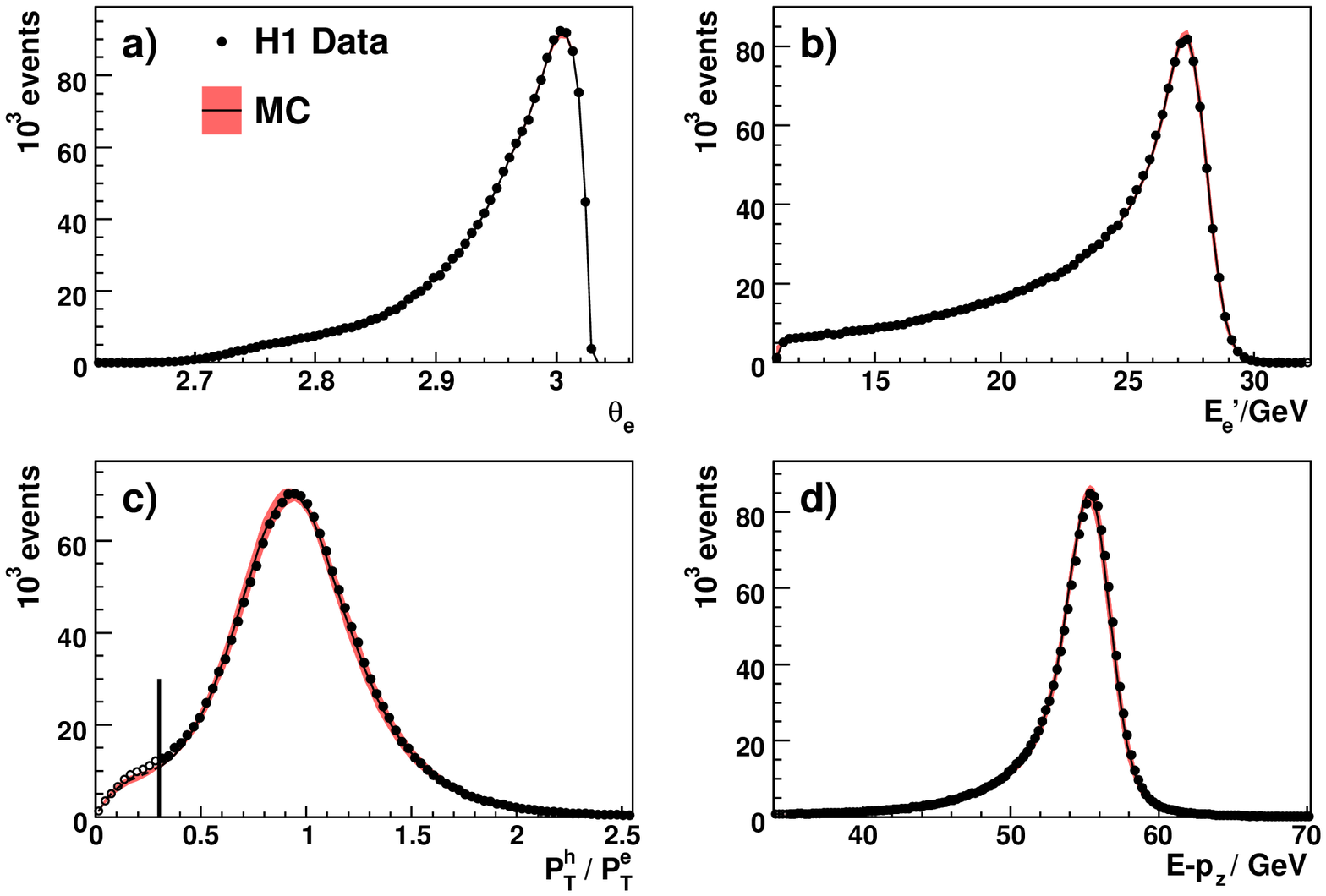}{.8\linewidth}
{Control distributions of the polar angle a) and the energy b) of
  the scattered positron, the transverse momentum ratio $\pth/\pte$
  c) and $\empz$ d).}{}

The good control of the relevant detector quantities expresses itself
in small systematic uncertainties. For example, the positron energy
scale is controlled to better than $0.2-1\%$ and the HFS energy to $\sim
2\%$. The relevant detector and reconstruction efficiencies are known
to typically better than $0.3 - 0.5\%$.
The event kinematics is reconstructed using the Electron method at
larger $y \gtrsim 0.1$ and the $\Sigma$ method at lower $y$~\cite{h1lowq2}.

\section{Combination with Published Results and Results on $F_2$}

The new measurement~\cite{h1medq2} covers a similar kinematic domain as the
previously best H1 measurement using data from 1996/97 with a proton
beam energy of $820$\,GeV~\cite{h197}. The total uncertainties are
improved by up to a factor $\sim 2$ for newer data. The comparison
with the older data revealed a small $Q^2$ dependent bias in the
published results of $\sim 0 - 2.5\%$, which is corrected.

The two data sets are averaged using the same technique as introduced
for the combination of the lower $Q^2 \leq 12\gevsq$ data sets of
H1~\cite{h1lowq2} with a proper treatment of bin-to-bin correlated
uncertainties. The 1996/97 cross sections are adjusted for the small
difference in the centre of mass energies for lower $y < 0.35$.
Measurements at higher $y$ are kept separate. The correlated
systematic uncertainty sources for measurements within a data set are
treated as uncorrelated between the data sets, only small shifts of
the central values are required. The consistency between the data sets
is good with a $\chi^2_{\rm tot}/\dof = 51.6/61$.

The combined data represent the most precise inclusive measurement in
the covered kinematic domain to date. The structure function $F_2$ is
extracted for $y<0.6$ by applying a small correction for the $F_L$
influence, which is taken from the NLO QCD fit H1PDF~2009, see
section \ref{sec:qcdfit}. Figure \ref{fig:kretzschmar_jan.fig2.eps}
shows the evolution of H1 measurement uncertainties on the structure
function $F_2$ at $Q^2=15\gevsq$. The typical total uncertainties have
now reached a level of $1.3 - 2 \%$.

\stdfig{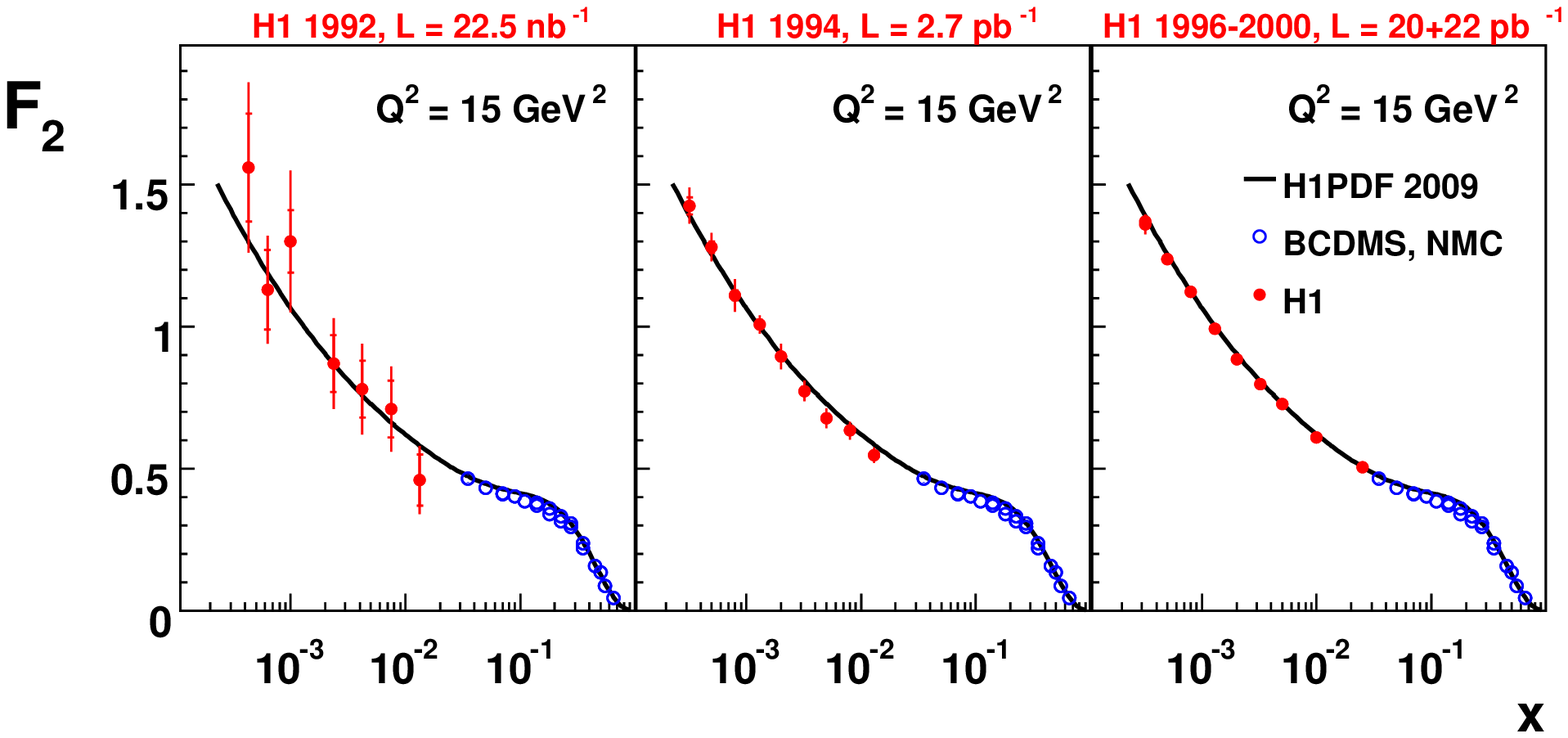}{.8\linewidth} {Measurements of the
  structure function $F_2$ at $Q^2 = 15\gevsq$ by the H1 collaboration
  at low $x$ complemented by fixed target data at higher $x$. The
  precision is seen to increase impressively over time, with the
  latest result reaching typically $1.3-2\%$ total uncertainty.}{}

The steep rise of $F_2$ towards low $x$ is well described by QCD fit.
At low $x < 0.01$ the rise of $F_2$ towards low $x$ essentially
compatible with a power law $F_2 \propto x^{-\lambda}$, but a small
deviation from this behaviour at lowest $x$ is also not excluded.
Figure \ref{fig:kretzschmar_jan.fig3.eps} shows a compilation of H1
data on $F_2$ at fixed $x$, including results from
lower~\cite{h1lowq2} and higher $Q^2$~\cite{h1pdf2000}. Strong scaling
violations are observed at low $x$, while at high $x\sim 0.1$ $F_2$ is
nearly independent of $Q^2$. The effect of the QCD dynamics
is well described by the fit down to very low $Q^2 \sim 1.5\gevsq$.

\stdfig{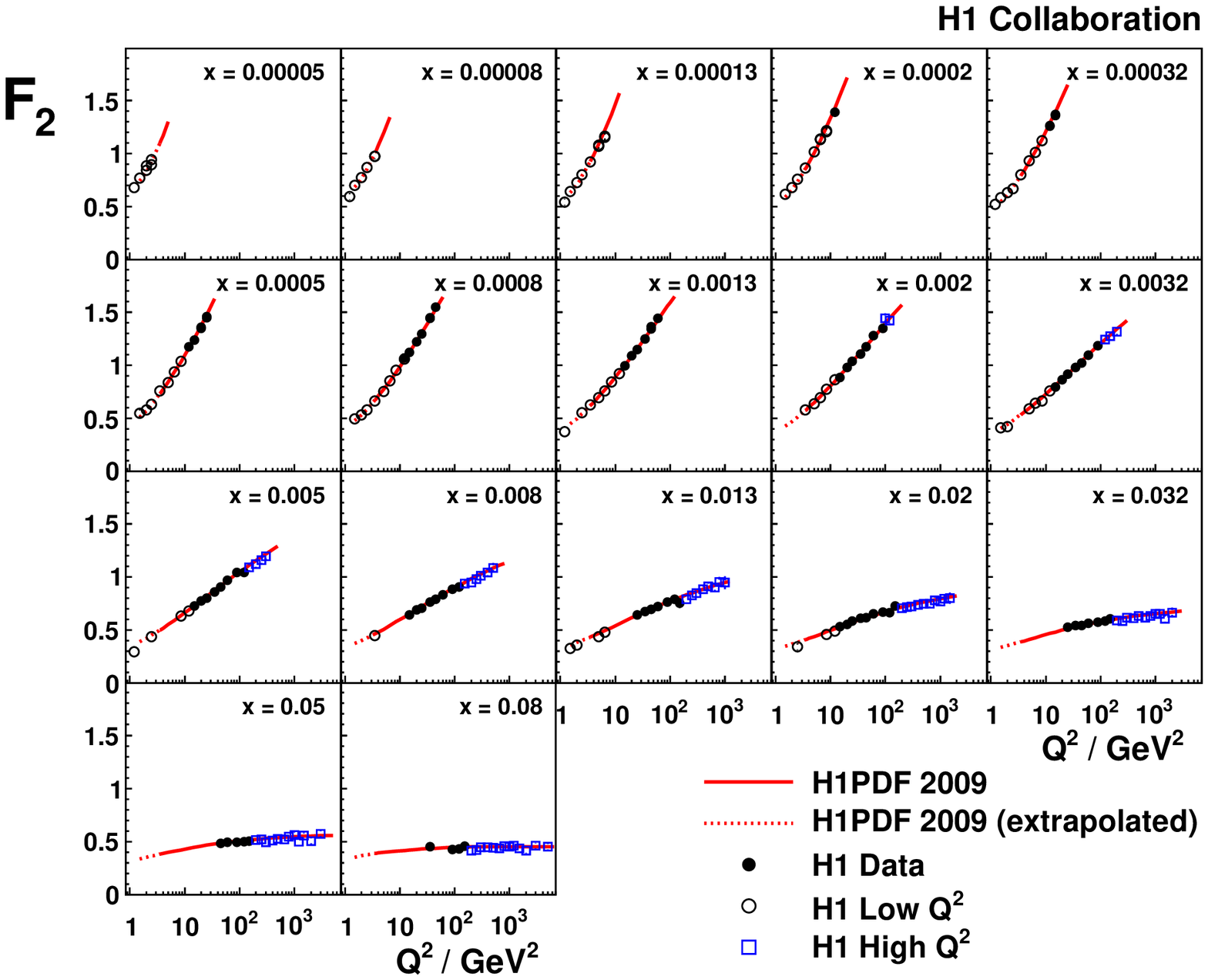}{.7\linewidth} {Measurements of the
  structure function $F_2$ by the H1 collaboration at constant $x$ as
  a function of $Q^2$.}{}

\section{The H1PDF 2009 QCD Fit}
\label{sec:qcdfit}

Using only inclusive cross section data by H1, a new QCD fit based on
DGLAP evolution equations at NLO was performed. Compared to the
previously published fit H1PDF~2000~\cite{h1pdf2000}, the data used at
lower $Q^2 \leq 150\gevsq$ includes two new
analyses~\cite{h1medq2,h1lowq2} and is more precise. Also the
theoretical treatment of heavy quark threshold effects was improved by
employing a GM-VFNS scheme~\cite{mstw2008}.

The chosen PDF set $xu_v, xd_v, xg, x\bar{U} = x(\bar{u}+\bar{c}),
x\bar{D} = x(\bar{d}+\bar{s}+\bar{b})$ is parameterised at low $Q^2_0
= 1.9\gevsq$ as $xP = Ax^B(1-x)^C(1+Dx+Ex^2+\ldots)$. Six parameters
are fixed by model assumptions and momentum and quark counting rules.
The remaining nine $A, B, C$ parameters form the basic
parameterisation. Further $D, E, \ldots$ parameters are considered
in an iterative $\chi^2$ optimisation procedure. Additional conditions
apply for parameters, which are to be used for the fit eventually.
Both structure functions $F_2$ and $F_L$ should be positive.
Furthermore, if fits with negative PDFs or very low valence quark
compared to the sea quark distributions at high $x$ are found, those
are not considered for the central result, but used to give an
estimate of the parameterisation uncertainty. Eventually only one
parameter, $D_g$, is added to the basic parametrisation.
This fit has a good $\chi^2_{\rm tot}/\dof = 587/644$ with no
significant tension in the systematic uncertainties.

The value of the strong coupling constant is fixed to $\alpha_s =
0.1176$. The further theory model parameters are chosen as $Q^2_0 =
1.9\gevsq$, $m_c = 1.4\,\mathrm{GeV}$, $m_b = 4.75\,\mathrm{GeV}$ and
the strange sea fraction $f_s=x\bar{s}/x\bar{D} = 0.31$, which are varied in
reasonable limits~\cite{h1medq2}. The
largest model uncertainty occurs on the gluon density at low $x$,
which is caused by the $Q^2_0$ variation.

\begin{figure}[tbh]
  \centering
  \includegraphics[width=0.32\linewidth]{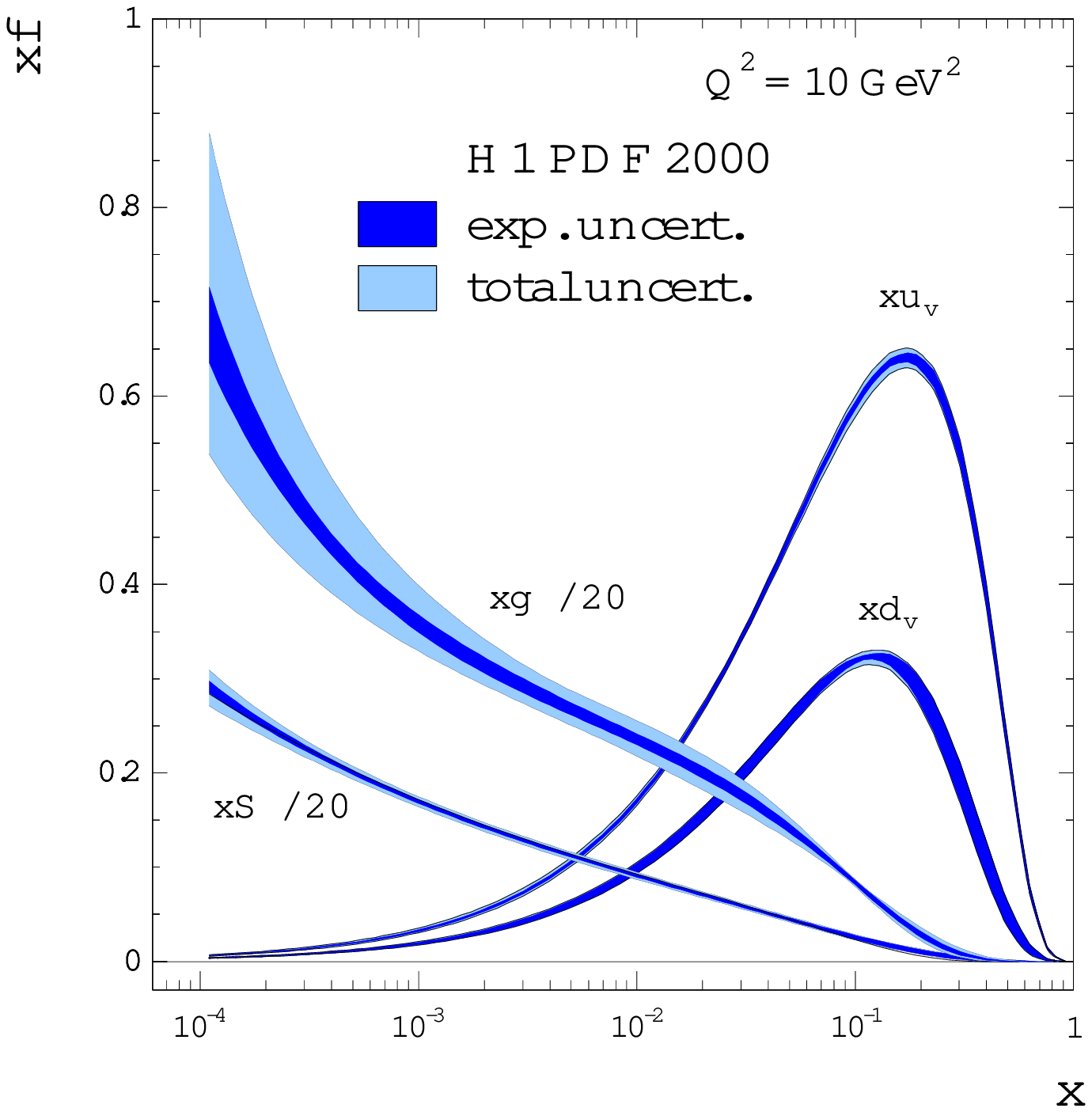}%
  \includegraphics[width=0.34\linewidth]{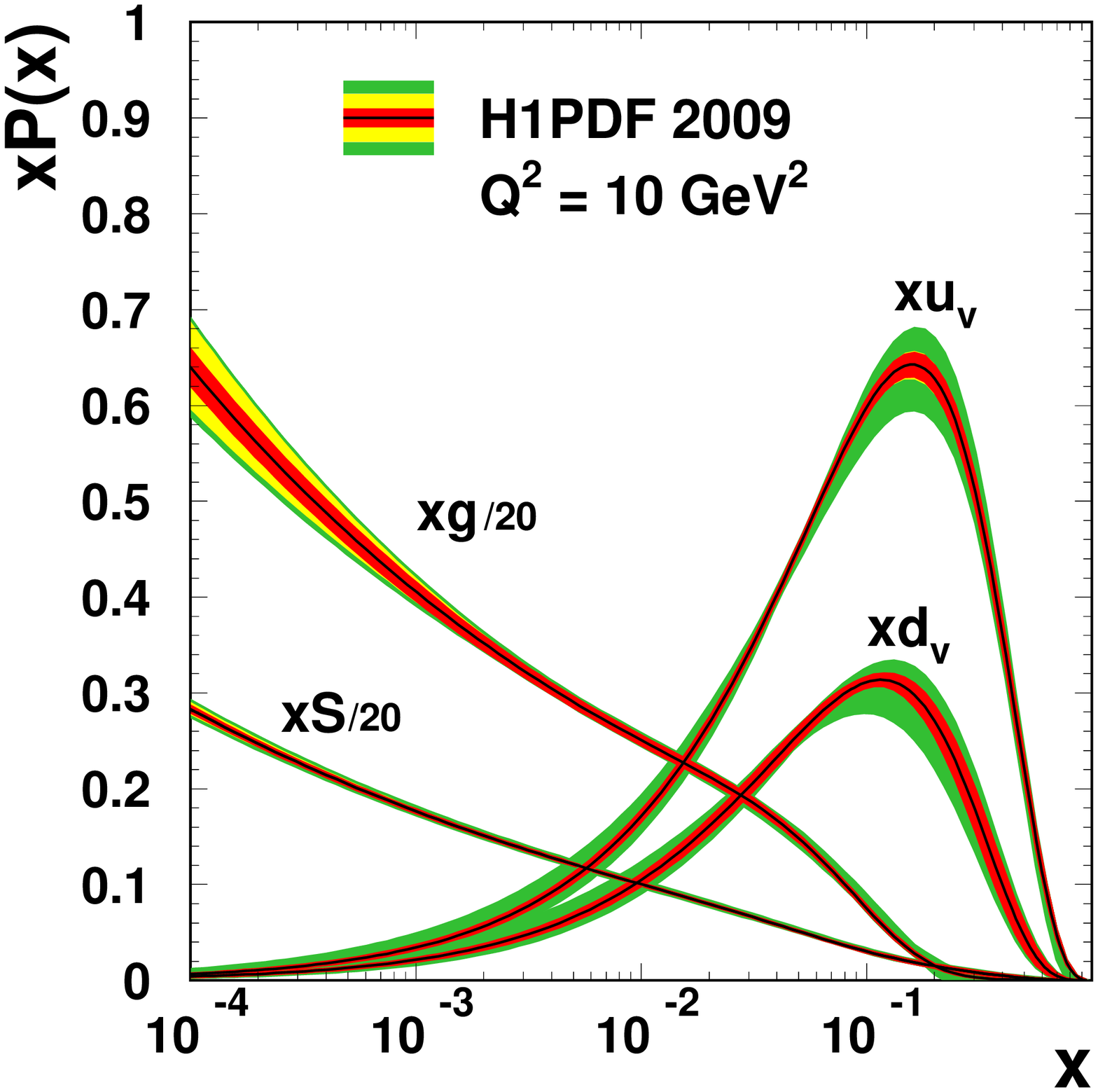}%
  \includegraphics[width=0.34\linewidth]{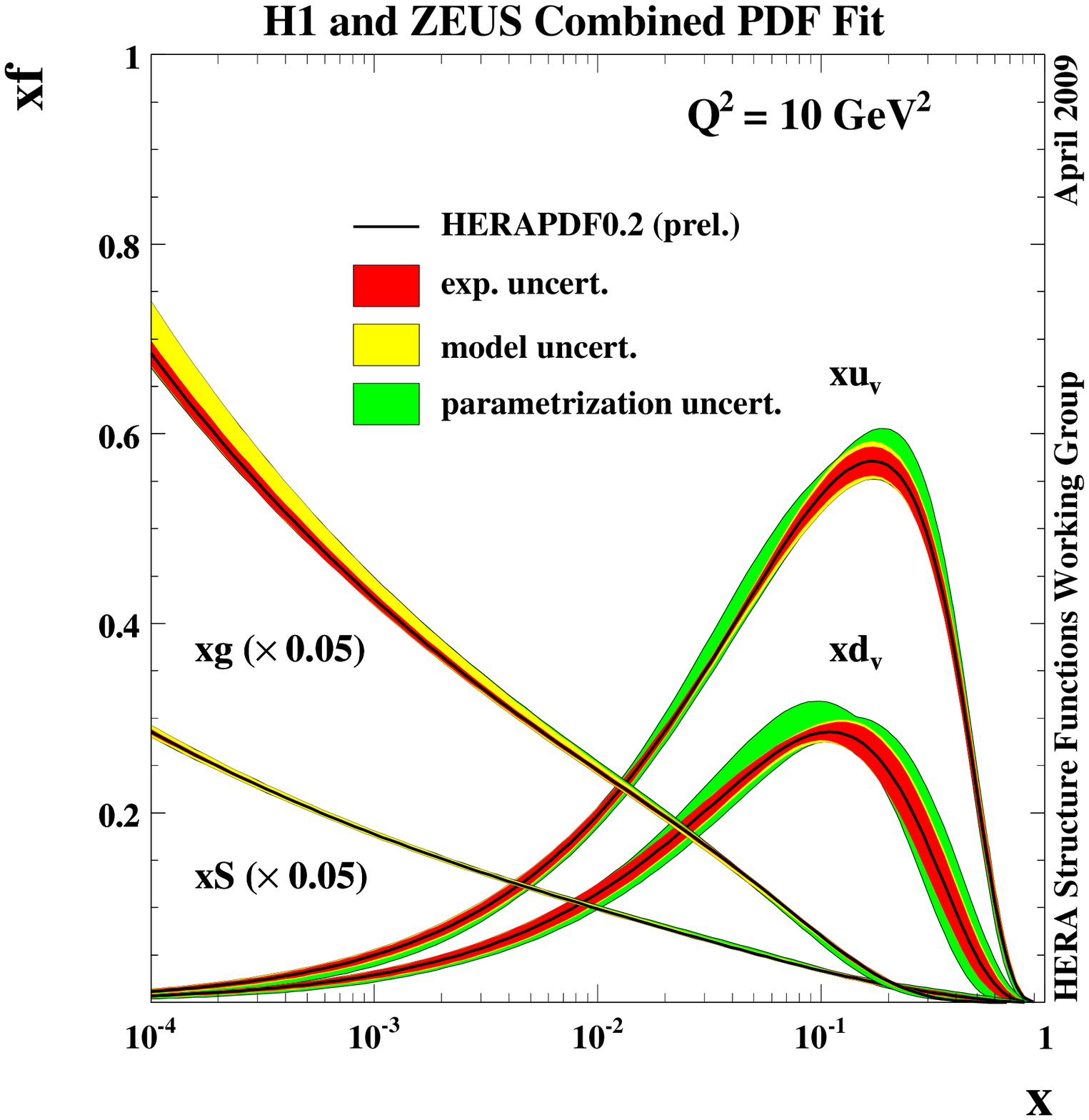}
  \caption{PDFs of the previous and current QCD fits by the H1
    collaboration (left and middle) and the preliminary combined H1
    and ZEUS fit (right) at $Q^2=10\gevsq$. Gluon and sea quark
    distributions are scaled down by a factor of $20$. The bands denote the
    experimental, model, and parametrisation uncertainties from inner
    to outer band.}
  \label{fig:fits}
\end{figure}

The PDFs for the valence quarks, the gluon, and the total sea density
$xS=x(\bar{U}+\bar{D})$ are shown with all uncertainties in figure
\ref{fig:fits} at a scale of $Q^2=10\gevsq$. The experimental
errors are in general small compared to the model and
parametrisation errors. The gluon clearly dominates low $x$
domain even at this relatively low scale. Compared to H1PDF~2000 the
low $x$ uncertainties are reduced, while the uncertainties at high $x$
larger and more realistic. Comparing to the latest combined H1 and
ZEUS fit, HERAPDF $0.2$~\cite{h1zeuscomb}, similar features are
observed. The combined fit profits from the data combination,
which is seen to constrain the PDFs further especially at higher $x$.

\section{Conclusions} 
A new measurement of the inclusive DIS cross section for
$12\gevsq \leq Q^2 \leq 150\gevsq$ is performed using H1 data from the
year 2000. The new measurement is combined with the published 1996/97
data, after correction of a small bias in the older result. The result
is the most accurate measurement in this kinematic domain to date with
typical total uncertainties of $1.3 - 2\%$. A QCD analysis, H1PDF~2009,
is performed using improved new data at $Q^2\leq 150\gevsq$ and
published H1 high $Q^2$ inclusive cross section measurements. The fit
is able to describe all the data very well and supersedes the
previous H1 fit.



\begin{footnotesize}

\end{footnotesize}



\begin{thebibliography}{99}
\bibitem{url} Slides: \\ 
  \verb$http://indico.cern.ch/materialDisplay.py?contribId=56&sessionId=0&materialId=slides&confId=53294$

\bibitem{h1medq2}
  F.D. Aaron {\it et al.} [H1 Collaboration],
  arXiv:0904.3513 [hep-ex].

\bibitem{h1zeusfl}
  F.~D.~Aaron {\it et al.}  [H1 Collaboration],
  Phys.\ Lett.\  B {\bf 665} (2008) 139
  [arXiv:0805.2809 [hep-ex]];\\
  A.~Glazov [H1 Collaboration], presented at DIS 2009, H1prelim-09-044;\\
  S. Chekanov {\it et al.} [ZEUS Collaboration],
  [arXiv:0904.1092 [hep-ex]].

\bibitem{h1lowq2}
  F.D. Aaron {\it et al.} [H1 Collaboration],
  arXiv:0904.0929 [hep-ex].

\bibitem{h197}
  C.~Adloff {\it et al.}  [H1 Collaboration],
  Eur.\ Phys.\ J.\ C {\bf 21} (2001) 33
  [arXiv:hep-ex/0012053].

\bibitem{h1pdf2000}
  C.~Adloff {\it et al.}  [H1 Collaboration],
  Eur.\ Phys.\ J.\ C {\bf 30} (2003) 1
  [arXiv:hep-ex/0304003].

\bibitem{mstw2008}
  A.~D.~Martin, W.~J.~Stirling, R.~S.~Thorne and G.~Watt,
  arXiv:0901.0002 [hep-ph].

\bibitem{h1zeuscomb}
  V. Racescu [H1 and ZEUS Collaborations],
   presented at DIS 2009, H1prelim-09-045, ZEUS-prel-09-011.

\end{thebibliography}
\end{document}